\documentclass[RNAAS]{aastex63}

\graphicspath{{./}{figures/}}

\begin{document}

\title{Origin of sources of repeating fast radio bursts with periodicity in close binary systems}

\correspondingauthor{Sergei B. Popov}
\email{sergepolar@gmail.com}


\author[0000-0002-4292-8638]{Sergei B. Popov}
\affiliation{Lomonosov Moscow State University, Sternberg Astronomical Institute.
Universitetski pr. 13, Moscow 119234 Russia}
\affiliation{National Research University ``Higher School of Economics'', Department of Physics.
Myasnitskaya str. 20, Moscow 101000, Russia}

\keywords{stars: neutron --- stars: magnetars --- stars: binary --- magnetic fields}

\begin{abstract}
It is proposed that sources of repeating fast radio bursts with detected periodicity are magnetars with extremely  short initial spin periods at the protoneutron star stage, formed in binaries after tidal synchronization of their progenitor cores on late stages of thermonuclear burning in very close systems with orbital periods about a few days. This formation channel is in several respects different from evolution of progenitors of normal magnetars, and thus, it explains some differences between repeating and non-repeating sources of FRBs.
\end{abstract}

\section{FRBs and magnetars} 

Fast radio bursts (FRBs) are enigmatic extragalactic transient radio sources discovered in 2007 and actively studied since 2013, see reviews in \cite{2018PhyU...61..965P, 2019A&ARv..27....4P, 2019ARA&A..57..417C}. 

Many models explaining properties of FRBs were proposed, see the list in \cite{2019PhR...821....1P}. Presently, the leading approach is related to a rapid release of magnetic field energy of magnetars \citep{2007arXiv0710.2006P, 2019arXiv190807743B, 2020ApJ...889..135L}. 
This line of consideration was recently supported by simultaneous discovery of a radio and X-ray flare from a known Galactic magnetar \citep{2020arXiv200510324T, 2020arXiv200510828B, 2020arXiv200506335M, 2020arXiv200511178R, 2020arXiv200511071L}.

Among FRBs there is a small group of repeating sources. Only two of them show numerous bursts, which finally allowed periodicity to be found: 16.35 days in FRB 180916.J0158+65 \citep{2020arXiv200110275T} and $\sim$157 days in FRB 121102 \citep{2020MNRAS.tmp.1508R}. The nature of this periodicity is still debated.

These prolific repeaters might either represent a subsample of extreme magnetars, or be very young objects in comparison with others. It was proposed that extreme properties of repeating FRBs can result from 
coalescence of two compact objects (see  references in \citealt{2019PhR...821....1P}). 
However, if periodicity is described in a binary system scenario \citep{2020ApJ...893L..39L, 2020ApJ...893L..26I} these variants are excluded (due to relatively short periods coalescence of internal binary in a triple system can be also excluded), as well as coalescence of normal stars at the common envelope stage which also produces rapidly rotating neutron stars (NSs) with extreme parameters \citep{2016MNRAS.463.1642P}. Then it is necessary to figure out a scenario of extreme magnetar formation in a  binary which can survive all preceding stages and the supernova explosion.

The fact that both prolific repeaters are found in binaries is puzzling, as the fraction of NSs in binaries is below 20\% \citep{2019MNRAS.485.5394K}. In addition, all known Galactic magnetars (soft gamma-ray repeaters and anomalous X-ray pulsars) are isolated objects \citep{2015RPPh...78k6901T}.

Thus, here we face a problem. It is necessary to bring in correspondence three items: hypothesis of binarity of all known repeating periodic FRBs, the fact of low fraction of NSs in survived binaries, and 100\% isolated nature of Galactic magnetars. A natural approach to solve this problem is to identify a unique framework of magnetar formation in the course of stellar evolution in which, on one hand, majority of magnetars have binarity fraction lower than typical NSs, but on other hand, extreme objects mainly are members of binary systems.
Here I briefly present a concept of such scenario based on our earlier studies of magnetar formation.

\section{Origin of magnetars related to sources of repeating FRBs with periodicities}

In this note I follow the line reviewed in \cite{2016A&AT...29..183P}:  high field of a magnetar is due to dynamo mechanism active in a rapidly rotating protoneutron star \citep{1993ApJ...408..194T}, rapid rotation of the progenitor core is due to interactions in a binary system \citep{2006MNRAS.367..732P, 2009ARep...53..325B}.

In \citep{2006MNRAS.367..732P} the authors demonstrated that two evolutionary channels can potentially explain the fraction of magnetars among all NSs and simultaneously nearly zero fraction of magnetars in binaries. These are: coalescence of stars prior to core collapse and  spin-up of the secondary companion (which produces a magnetar) due to accretion from the primary one. \cite{2009ARep...53..325B} analysed a different channel: tidal synchronization of the progenitors' core on late stages of thermonuclear burning. In this case, a magnetar progenitor explodes in a  very tight binary with orbital period $\sim$ few days. Stellar cores spun-up in this evolutionary channel are unable to loose their angular momentum via stellar winds or envelope ejection. Unless the NS gets a huge kick ($\gtrsim$500-700 km s$^{-1}$), the system survives. The formation rate through this channel is lower that the magnetar formation rate, especially if synchronization is reached on very late stages of stellar evolution, which potentially might lead to formation of more rapidly rotating protoNSs.

Here I propose a hypothesis, that majority of FRBs are due to magnetars formed through more probable channels of binary evolution, and prolific repeaters  are extreme magnetars formed from rapidly rotating cores which have been tidally synchronized on very late stages of the progenitor evolution. 

In order to spin-up the core due to tidal synchronization it is necessary to have a strong core-envelope coupling \citep{2016MNRAS.463.1642P}. Potentially, black holes also can form from such rapidly rotating cores \citep{2019MNRAS.483.3288P}.
In this respect, such extreme magnetars are ``cousins'' of gamma-ray burst (GRB) progenitors. I.e., in the same channel a rapidly rotating black hole can be produced, and then after its formation it can appear as a long GRB. The fraction of GRBs in comparison with all core-collapse supernova is about 3\% \citep{2007AIPC..937..492S}. This number is likely close to the expected fraction of extreme magnetars among all NSs. The fraction of all magnetars is sensibly larger \citep{2019MNRAS.487.1426B}, more in correspondence with modeling by \cite{2006MNRAS.367..732P}. Note, however, that the birth rate of magnetars which have progenitors synchronized just before collapse is low --- $\sim 1/100 000$~yrs, --- according to \cite{2009ARep...53..325B}.  

Post-explosion and pre-explosion orbital parameters are linked as \citep{2014LRR....17....3P}:

\begin{equation}
    \frac{a_\mathrm{f}}{a_\mathrm{i}}=\left[ 2-\chi\left( \frac{w_\mathrm{x}^2+(V_\mathrm{i}^2+w_y^2)+w_\mathrm{z}^2}{V_\mathrm{i}^2}\right)  \right]^{-1}.
\end{equation}
Here, $a_\mathrm{f}$ and $a_\mathrm{i}$ are final and initial semi-major axes, $V_\mathrm{i}$ is the initial (orbital) velocity, $w$ is the kick velocity, and $\chi=(M_1+M_2)/(M_\mathrm{c}+M_2)$ with $M_1, M_2$ --- masses of components before the explosion, and $M_\mathrm{c}$ --- the NS mass.
It is quite easy to obtain orbital periods  $\sim 10-100$ days for moderate kick values and mass loss (note, that before the explosion the star might already loose its outer envelopes due to interactions in the binary). Formation rate of such systems, as demonstrated by population modeling \citep{2009ARep...53..325B}, is non-negligible, but significantly smaller than for normal magnetars.

In some cases due to large kicks the system with a magnetar from a tidally synchronized core can be destroyed, and so it is possible to obtain a small number of isolated repeating FRB sources which would not then demonstrate any periodicity.

Most probable companions of extreme magnetars, according to modeling by \cite{2009ARep...53..325B}, are main-sequence stars (if it is the first explosion in the system), or black holes (if this is the second explosion). However, in scenarios like those proposed by \cite{2020ApJ...893L..39L, 2020ApJ...893L..26I} a black hole cannot be responsible for any periodicity in FRBs appearance. Thus, we are rested with  main sequence stars as typical companions of the sources of interest. As usually  stars in a binary system have a tendency to be of comparable masses, we can expect that typically companions of extreme magnetars have masses of NS progenitors.

\acknowledgments
The author acknowledges the support from the Program of development of M.V.Lomonosov Moscow State University (Leading Scientific School ``Physics of stars, relativistic objects, and galaxies''). The author thanks Dr. M Barkov and Prof. K. Postnov for discussions and comments.

\bibliographystyle{aasjournal} 
\bibliography{frb}

\end{document}